# Simple approach for detection and estimation of photoactivity of silver particles on graphene oxide in aqueous-organic dispersion

**D.V.Vlasov, L.A. Apresyan, T.V.Vlasova**

AM Prokhorov General Physics Institute
Moscow, Russia

ABSTRACT

In the aqueous-organic (containing dimethylformamide) dispersion of graphene oxide flakes with deposited thereon Ag-particles, under the action of light in visible range, apparently for the first time, the effect of sediment flotation was observed with subsequent stabilization of the dispersion, which does not occur in the absence of Ag-particles. The main reason for such a laser light induced movement of sediment GO flakes may be explained with the appearance of small bubbles. The further development of this approach seem to be able to estimate photo activity of graphene flakes with different activation particles.

Оксид графена (GO) и восстановленный оксид графена (RGO) в последние годы привлекают все возрастающее внимание исследователей, поскольку позволяют реализовать большую часть преимуществ двумерного графена, даже в тех случаях, когда они востребованы на микроуровне, т.е. на уровне пластинок однослойного графена микронных размеров [1-4]. При этом в случае использования GO или RGO удается реализовать воспроизводимые характеристики электронных устройств, а в ближайшей перспективе, и промышленные «дешевые» технологии производства, которые пока недоступны для CVD или PECVD методов синтеза малослойных и однослойных листов графена.

Одно из важных направлений RGO-электроники связано с развитием солнечной энергетики, направленной на создание фотоэлектрических преобразователей, а также фотокаталитический синтез экологически чистого водородного топлива [5,6]. Для этого, в частности, на GO высаживают частицы светочувствительного материала (поглощающего с образованием и разделением зарядов) например $TiO_2$ [7,8], ZnO [9], CdS [10], наночастиц Au [11], Ag [12] и т.д.

В настоящей заметке в водно-органической (содержащей диметилформамид, ДМФА) дисперсии хлопьев оксида графена с осажденными на них частицами серебра под воздействием света в видимом диапазоне, по-видимому впервые, наблюдался эффект всплывания осадка с последующей стабилизацией дисперсии. В отсутствие частиц Ag указанный эффект не проявляется.

Для проведения этих экспериментов из мелкодисперсного графита (фирма Asbury, Carbon, lot 9437) по стандартной процедуре Хаммера [13] (обработка смесью минеральных кислот с перманганатом калия) приготавливали GO, затем после обработки ультразвуком раствор разделяли на две части: эталонную GO и часть для экспериментов по осаждению частиц серебра. В экспериментальной взвеси GO используя методику высаживания наночастиц серебра [14], добавляли раствор нитрата серебра 1 мг на 1 мл, ДМФА 20 мл. и на магнитной мешалке перемешивали раствор в течение шести часов при $60^0$ С. Раствор при этом изменял цвет с желто-коричневого до темно-коричневого, отстаивался в течение суток, после чего возникло расслоение: сверху слой «чистого» прозрачного раствора, не содержащего хлопья GO с частицами серебра и придонный осадок из серо-коричневых хлопьев (Рис.1.)

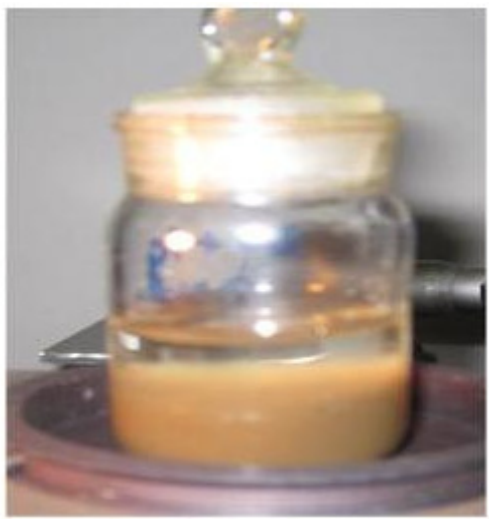

Рис.1 Расслоившийся раствор GO с осажденными частицами Ag.

Собственно наличие частиц серебра контролировалось оптическим микроскопом, где наблюдались металлические частицы с максимальным размером до 5 мкм. Далее для исследования полученного композита он облучался сверху непрерывным лазером 532 нм с максимальной мощностью 100 мВт. При облучении лазером наблюдалось движение осажденных частиц вверх, в направлении «навстречу» лазерному излучению. В результате вблизи лазерного луча образовывалась «выброс» осадка, далее выброс дорастал до поверхности жидкости. И долго (десятки минут) сохранялся практически в неизменном виде, скорее происходил развал выброса и взвешенные частицы расплывались по всему прозрачному объему. За сутки все частицы вновь успевали осесть.

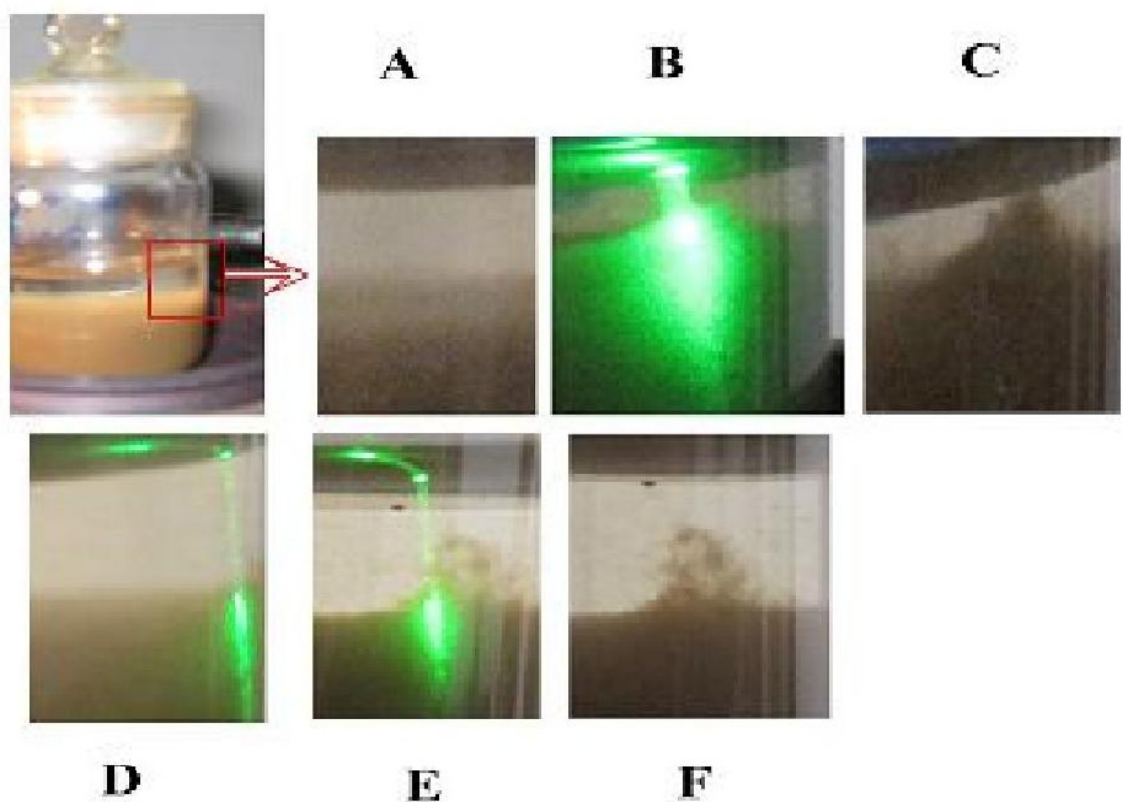

Рис.2 Эксперименты по лазерному втягиванию хлопьев GO с частицами серебра. В верхнем левом углу фото исходной кюветы до начала экспериментов, квадратиком выделена область, в которой с увеличением отслеживается втягивание частиц в лазерный A>B>C фотографии участка выделенного красным квадратом на исходной кювете: A- до освещения лазером, B — в процессе образования выброса, C — после выключения лазера. Фотографии D>E>F то же самое, только для сфокусированного луча.

Аналогичный эффект обнаружился и при воздействии на полученную водно-органическую дисперсию хлопьев графена с частицами серебра солнечного излучения. При этом наблюдалось изменение цвета коллоида до почти черного и всплывание всей облученной солнечным светом массы RGO с частицами Ag (Рис.3).

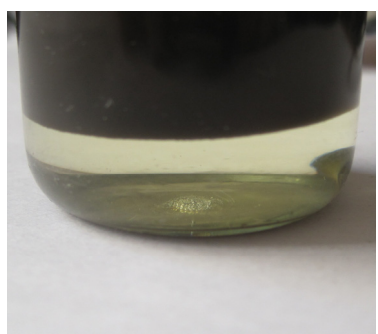

Рис.3. Всплывание коллоида RGO с частицами Ag после воздействия солнечным излучением в течение 8 часов.

Наблюдаемое изменение цвета коллоидных частиц можно объяснить восстановлением под действием света GO до RGO, которое обусловлено фотокаталитической активностью частиц серебра, описанному в работе [15] для случая УФ облучения.

Основной механизм перемещения взвешенных частиц, в том числе коагуляции коллоидных растворов согласно современным представлениям связан с возникновением двойного зарядового слоя [16], образующегося при нормальных значениях PH во взвешенных частицах, которые осаждаются при его увеличении. Однако этот механизм не позволяет объяснить наблюдающееся всплывание коллоида под воздействием излучения. Такое всплывание можно связать с возникновением обеспечивающих положительную плавучесть частиц мельчайших пузырьков газов в результате фотокаталитического разложения воды, либо дегидрированием имеющегося на RGO остатков водорода, аналогичного рассмотренному в недавней работе [17]. Тем более, что при перемешивании всплывшей взвеси с помощью ультразвука и получении полностью однородной дисперсии, расслоение вверх под действием солнечного излучения воспроизводилось в течение нескольких часов.

Дальнейшее изучение описанных явлений может представлять интерес в связи с потенциальной возможностью создания элементов искусственного фотосинтеза или полимерно-композитных солнечных батарей, причем сам эффект «лазерного всплывания» , по-видимому, может быть использован для определения фото-электро-активности частиц графена с различными добавками, активизирующими процесс фотоэлектролиза. Отметим также, что все эксперименты проводились параллельно и одновременно с дублирующим раствором взвеси хлопьев GO без металлических частиц, причем никаких движений подобных описанным выше, в этих взвесях не наблюдалось.